\documentstyle[preprint,aps]{revtex}
\draft
\title
{Wei Hua's Four-Parameter Potential \\ Comments \\ and Computation of
Molecular Constants \\ $\alpha_e$ and $\omega_e x_e$}
\author
{Sarvpreet Kaur and C. G. Mahajan\thanks{e-mail: spectphy@ch1.vsnl.net.in} }
\address
{Centre of Advanced Study in Physics,\\
Panjab University, Chandigarh - 160 014 (India).}
\date{\today}
\def\be{\begin{equation}}
\def\bea{\begin{eqnarray}}
\def\ee{\end{equation}}
\def\eea{\end{eqnarray}}
\def\nn{\nonumber}
\def\bt{\begin{table}}
\def\btab{\begin{tabular}}
\def\et{\end{table}}
\def\etab{\end{tabular}}
\begin{document}
\maketitle
\begin{abstract}
The value of adjustable parameter $C$ and the four-parameter potential
$U(r) = D_{e}\left [ \frac{1-\mbox{exp}[-b(r-r_{e})]}{1-C\mbox{exp}
[-b(r-r_{e})]} \right ]^{2}$
has been expressed in terms of molecular parameters and its
significance has been brought out. The potential so constructed,
with $C$ derived from the molecular parameters, has been applied
to ten electronic states in addition to the states studied by
Wei Hua. Average mean deviation has been found to be $3.47$
as compared to $6.93$, $6.95$ and $9.72$ obtained from Levine2,
Varshni and Morse potentials, respectively. Also Dunham's
method has been used to express rotation-vibration interaction constant
$(\alpha_{e})$ and anharmonocity constant $(\omega_{e}x_{e})$ in terms of
$C$ and other molecular constants. These relations have been employed to
determine these quantities for $37$ electronic states. For $\alpha_{e}$,
the average mean deviation is $7.2\%$ compared to $19.7\%$ for Lippincott's
potential which is known to be the best to predict the values.
Average mean deviation for $(\omega_{e}x_{e})$ turns out to
be $17.4\%$ which is almost the  same  as  found  from Lippincott's
potential function.
\end{abstract}
\narrowtext
\section{Introduction}
A knowledge of exact potential function which governs the interaction between
atoms as a function of internuclear distance, is of fundamental importance in
a wide variety of fields ranging from gas kinetics to stellar structure.
Besides, potential energy (PE) curves provide a good deal of information about
the molecular structure. Accordingly, a number of methods have been developed
to obtain these curves and the most satisfactory technique for this purpose
is RKR method \cite{r1,r2,r3}, which is based on the experimental spectroscopic
data. This approach is, however, limited to the region for which the
spectroscopic data exist. To overcome this limitation, one resorts to either
Dunham method or to analytical functions. In the former, term values are
expressed as:
\be
T = \sum_{i,j}Y_{i,j}(v+\frac{1}{2})^{i}J^{j}(J+1)^{j} ,
                                                               \label{1}
\ee
where v and J are the vibrational and rotational quantum numbers, respectively,
and the coefficients $Y_{i,j}$ are related to spectroscopic constants.
The formulation of analytical functions is based on the assumption
that bonding potential curves can be fitted to a certain form of
algebraic expression. A comparative study of the empirical potential
functions by Steele et al \cite{r4} and by Varshni \cite{r5}
revealed that of all the three-parameter potential functions
those suggested by Morse \cite{r7}, Varshni \cite{r5} and
Levine \cite{r8} have small average absolute deviation from the
RKR curve with their respective values as $3.68\%$, $2.31\%$
and $1.98\%$. The corresponding Schrodinger equation can, however, be exactly
solved only for Morse function.
\par
Some time back Wei Hua \cite{r9}, introduced a four-parameter (FP) potential
function which apart from the usual three parameters viz. $\omega_{e}$,
$r_{e}$, $D_{e}$, contains a fourth parameter, b, which in turn, depends
on an adjustable constant $C$. The value of $C$ was chosen to minimize the
absolute deviation of the calculated values from the corresponding
RKR - values. Wei Hua studied $15$ electronic states and found that
FP curves fit the RKR curves more closely compared to Morse curves.
Moreover, the corresponding Schrodinger equation can be solved exactly
for zero and approximately for non-zero total angular momentum \cite{r9}.
Almost all the researchers who have put forward an empirical potential
function including the FP potential or have performed the comparative studies,
have confined themselves to the study of simple molecules $\hbox{H}_{2}$,
$\hbox{O}_{2}$, $\hbox{N}_{2}$, CO, NO, hydrogen and alkali halides or to
alkali oxides. With a view to test the applicability of FP potential
to a wider range of molecules, Morsagh \cite{r10} carried out their comparative
study for diatomic molecules containing sulphur. It was found that for outer
wall of the potential curve $(r > r_{e})$ the average mean absolute deviations
from RKR curve for Morse, Varshni, Levine and four-parameter potentials
are $1.34\%$, $2.09\%$, $2.67\%$ and $0.7\%$, respectively. Encouraged by
the results in respect of FP potential, we have now dwelt upon the
significance of adjustable constant $C$ and have assessed how faithfully
the potential function predicts the values of molecular parameters
$\alpha_{e}$ and $\omega_{e}x_{e}$. This communication is an outcome of
these efforts.
\section{The  RKR-method}
This method \cite{r1,r2,r3} provides the classical turning points,
\bea
r_{max} &=& \left [ \frac{f}{g} + f^{2} \right ]^{\frac{1}{2}}+f , \nn \\
r_{min} &=& \left [ \frac{f}{g} + f^{2} \right ]^{\frac{1}{2}}-f
\label{2}
\eea
where f and g depend on the experimentally determined molecular
constants $(\omega_{e},\omega_{e}x_{e},B_{e},\alpha_{e})$ and are defined as
\bea
f &=& \frac{\partial{S}}{\partial{U}}, \nn \\
g &=& -\frac{\partial{S}}{\partial{k}}
                                                            \label{3}
\eea
with
\be
S(U,k) =
\frac{1}{\pi(2\mu)^{\frac{1}{2}}} \int_{0}^{I^{ \prime}} \{ U - E(I,k)\}^{
\frac{1}{2}}dI
                                                             \label{4}
\ee
Here U is the potential energy and E(I,k) is the sum of vibrational and
rotational energies of the molecule,
\bea
I &=& h(v+\frac{1}{2}),  \nn \\
k &=& \frac{J(J+1) h^2 } {8\pi^{2} \mu}
                                                                \label{5}
\eea
$\mu$ is the reduced mass of the molecule .
However, for the present work RKR potentials have been taken from the
literature and the references are given in tables 1 and 3.
\par
  Accuracy of a potential function can be judged in two ways: first by its
comparison with the curve obtained from the experimental data and second,
by using it to evaluate unused constants and comparing these with the
respective experimental values. In the reported work the findings have
been subjected to both the tests.
\section{Potential  Functions  for  comparative  study}
In the present work we have considered only those potential functions for
comparative study, which fit more closely to the RKR potential curve \cite{r4}.
The expressions for these potential functions and parameters $\alpha_{e}$
and $\omega_{e}x_{e}$ are as following:
\newline
{\bf (i) Morse potential \cite{r4,r7} }
\be
U_{M}(r) = D_{e}\left[ 1- e^{-a( r - r_e )} \right]^2 ,
                                                                \label{6}
\ee
\bea
\alpha_{e} &=& \left( \frac{6B_{e}^{2}} {\omega_{e} } \right) \left(
\Delta^{\frac{1}{2}} - 1 \right), \nn \\
\omega_{e}x_{e} &=& \left( \frac{\omega_{e}^{2}}{4D_{e}} \right);
                                                                \label{7}
\eea
{\bf (ii)   Varshni potential \cite{r5,r6}}
\be
U_{V}(r) = D_{e} \left[ 1 -
\frac{r_{e}} {r} e^{-b_{ \sc{v}}(r^{2}-r_{e}^{2} ) } \right]^{2},
                                                                \label{8}
\ee
\bea
\alpha_{e} &=& \left( \frac{6B_{e}^{2}} {\omega_{e} \Delta^{\frac{1}{2}}}
\right )\left ( \Delta-2\Delta^{\frac{1}{2}} +2 \right) , \nn \\
\omega_{e}x_{e} &=& \frac{B_{e}}{8} \left (
8\Delta-12+36-\frac{56}{\Delta^{\frac{1}{2}}}+\frac{48}{\Delta} \right);
                                                                 \label{9}
\eea
{\bf (iii)   Levine potential \cite{r4,r8}}
\be
U_{L}(r) = D_{e} \left[ 1 - \frac{r_{e}}{r}e^{ -b_{ \sc{L}} (r^{p} -
r_{e}^{p})} \right] ^{2},
                                                                  \label{10}
\ee
\bea
\alpha_{e} &=& \left( \frac{6B_{e}^{2}}{\omega_{e}} \right) \left(
\frac{3}{4}\Delta^{ \frac{1}{2}} - \frac{1}{2} \right), \nn \\
\omega_{e}x_{e} &=& \frac{B_{e}}{8}\left (
8\Delta - 12 (p-1) \Delta^{\frac{1}{2}} + 8p^{2} + 4 - \left[ \frac{ (20p^{2}
- 12p)}{\Delta^{\frac{1}{2}}} \right] + \frac{12p^{2}}{\Delta}\right);\nn \\
                                                                   \label{11}
\eea
{\bf (iv)    Four- parameter potential \cite{r9}}
\bea
U(r) &=& D_{e}\left[ \frac{1-\exp \{ -b (r-r_{e} ) \} } {1- C \exp
\{ -b (r - r_{e} ) \} } \right]^{2}
                                                                   \label{12}
\eea
The  expressions  for  $\alpha_{e}$ and $\omega_{e}x_{e}$ for this  potential
have  been  obtained  in  Section IV. \\
Here, $D_{e}$ is the dissociation energy, $r_{e}$ is the equilibrium
bond length and `a' is related to force constant through
\bea
k_{e} &=& 2D_{e}a^{2} = U'' (r_{e})   {\rm ~~~ with~~ } \nn \\
k_{e} &=&  \mu\omega^{2};
                                                         \label{13} \\
\Delta^{\frac{1}{2}} &=& ar_{e} ;                        \label{14} \\
b_{\sc{v}} &=& \frac{\frac{1}{2} \left( \Delta^{ \frac{1}{2}} - 1
\right) } {r_{e}^{2} };
                                                           \label{15} \\
b_{\sc{L}} &=& \frac{p^{-1} \left( \Delta^{\frac{1}{2}} - 1 \right)}
{r_{e}^{p}};
                                                             \label{16} \\
p &=& 2 + \frac{ \frac{1}{4} (\Delta^{\frac{1}{2}} - 4)
(\Delta^{\frac{1}{2}} - 2)} {(\Delta^{\frac{1}{2}}-1) }
                                                                \label{17}
\eea

In order to obtain an analytical expression for $C$, we have compared FP
potential with five-parameter potential function of Hulburt and
Hirschfelder \cite{r11} because a potential function with large number
of parameters is more flexible. This gives $C$ as
\be
C_{rel} = \left(- \frac{1}{12} + x \right)
 - \sqrt{ \frac{120}{144} - \frac{37}{48} x^{2} - \frac{2}{12}x +
 \frac{17}{144} \frac{G}{ar_{e}^{2} } }
                                                             \label{18}
\ee
where
\bea
x &=& \frac{1+F}{ar_{e}} , \nn \\
F &=& \frac{\alpha_{e}\omega_{e}}{6B_{e}^{2}},  \nn \\
G &=& \frac{8\omega_{e}x_{e}}{B_{e}} \nn
\eea
and
\bea
a &=& \sqrt{\frac{k_{\sc{e}}}{2D_{e}}}
                                                              \label{19}
\eea
$C$ has  been  denoted  as  $C_{rel}$  to  distinguish  it  from  the  $C$,
obtained  by  minimum
deviation  method  adopted  by  Wei  Hua.
\par
The values of $C_{rel}$ have been computed for the electronic states
studied by Wei Hua using the data listed in Table 1 and are compared
with $C$ (columns 3 and 4 in Table 2). Value of $C_{rel}$ depends on
$B_{e}$, $\alpha_{e}$ \mbox{\rm  and} $\omega_{e}x_{e}$, apart from other
parameters viz. $\omega_{e}$, $r_{e}$ \mbox{and} $D_{e}$, that are
used in almost all the three-parameter potential functions. In other
words, the value of $C_{rel}$ also depends on the rotational constants.
Furthermore, it is observed that a small change in $\alpha_{e}$ changes
$C_{rel}$ significantly while a similar change in other parameters produces
small variation. For example, in case of $Li_{2}$, if we change the value
of either $\alpha_{e}$ \rm or $\omega_{e}x_{e}$,  then resulting  variation
in  the  value  of $C_{rel}$  due  to  the  latter  is  nearly  one-third
that  due to the  former. Thus, $C_{rel}$ is  more  sensitive  to the
rotation-vibration  interaction  constant compared  to  other  parameters
involved  in  the  relation. Using  the equation  (18), values  of
$C_{rel}$  for  the  electronic  states   studied  by  Wei Hua,
have  been  calculated and  this, in  turn,  has  been  used  to
calculate  the  FP  potential  curve.
\par
Mean  square  deviations  in
\be
\Delta Y_{j} = \frac{\sqrt{ \left( \overline{U_{j}-U_{RKR} } \right)^{2} } }
{D_{e}}
                                                           \label{20}
\ee
where  $j$  denotes  four-parameter (FP),  Morse (M),  Levine (L),  Varshni (V)
potential functions,  have  been  calculated  for  all  the  cases.
The  results  are  projected  in  Table 2. It  has  been  found  that
the  mean  square  deviations  with  $C_{rel}$  are  not  far  off from the
corresponding  values  derived  with  $C$  which  makes  us  to have  faith
in  their correctness.  In  view  of  this,  $C_{rel}$ has  been  employed
to  calculate  the  potential  curve for  additional  ten  electronic states
(Table 3)  and  deduce  the  values  of  $\alpha_{e}$  and $\omega_{e}x_{e}$
for a total  of  37 electronic  states  of  various  molecules  including
the  ones  studied  by Wei Hua.
\section{Derivation of $\alpha_{\rm e }$
and  $\omega_{\rm e} x_{\rm e }$ }
On  the  basis  of  wave  mechanics  Dunham \cite{r12} has  shown
that molecular  parameters can  be  expressed  in  terms  of  the
derivatives  of potential  functions  if  it  can  be expanded  in
the   form  of  a  power  series. Applying  this  method, we  get
\be
\alpha_{e} = \left[ ar_{e}(1+C) - 1\right]\frac{6B_{e}^{2}}{\omega_{e}}
                                                              \label{21}
\ee
and
\bea
\omega_{e} x_{e} &=& \frac{h}{8\pi^{2}c\mu}a^{2}(1+C+C^{2}) \nn \\
                &=& B_{e}\Delta(1+C+C^{2})
                                                              \label{22}
\eea
 where $B_{e}=\frac{h}{8\pi^{2}cI}$ as obtained from the solution of equation
for  a  rotator. The relation  for  $\alpha_{e}$  is  of  the  same  form
as obtained by  Pekeris \cite{r12} using the Morse potential function, viz.
\be
\alpha_{e} = \left[ \left( \frac{\omega_{e}x_{e}}{B_{e}} \right)^{\frac{1}{2}}
- 1 \right]\frac{6B_{e}^{2}}{\omega_{e}} = \left[ ar_{e} - 1 \right]
\frac{6B_{e}^{2}}{\omega_{e}}
                                                              \label{23}
\ee
and  that  of  Lippincott's  relation \cite{r4} i.e.
\be
\alpha_{e} = \left( \frac{6B_{e}^{2}}{\omega_{e}}\right)
ab\Delta^{\frac{1}{2}}
                                                              \label{24}
\ee
where

\bea
a &=& \frac{4}{5} \left( 1 - \frac{1}{b\Delta^{\frac{1}{2}} } \right);
\nn \eea
and
\bea
b &=& 1.065
                                                               \label{25}
\eea
The  expression  for $\omega_{e}x_{e}$  is  to  be  compared  with
\be
\omega_{e}x_{e} = \frac{B_{e}}{8} \left( 3 + 12ab\Delta^{\frac{1}{2}}
+ 6\Delta + 15a^{2}b^{2}\Delta - 12ab^{2}\Delta \right)
                                                              \label{26}
\ee
obtained  from    Lippincott's  potential  function \cite{r4}.
\par
Values  of  these  two  parameters  have  been  calculated  for
$37$  electronic  states and are  given  in  Tables  4  and  5
for  $\alpha_{e}$  and $\omega_{e}x_{e}$, respectively. These  Tables
also include the  corresponding  experimental  values  and  the  values
found  by  employing Morse (eqn. 7), Varshni (eqn. 9) and  Lippincott's
(eqns. 23, 26) potential  functions. Values  of  molecular  constants
used  in  these  calculations  are  listed  in  Table  1.
\section{Results  and  Discussion}
The  adjusTable  constant  $C$    is  determined  by  finding  the mean square
deviation of  the  FP  potential  from the  RKR  potential  curve, which
itself  is  calculated  using  the molecular  parameters  $\omega_{e},
\omega_{e}x_{e}, \alpha_{e},r_{e}$ and $B_{e}$
of  the  individual  energy   levels \cite{r1,r2,r3}. Thus, any  variation
in  the  parameters  from  level  to  level  is  taken  care  of  and  hence
most  of  the  variations  are  absorbed  indirectly.  On  the  other  hand,
the  least  square fitted  values  of  the  parameters  are  used  in  the
expression  for  $C_{rel}$. These  values  of parameters,  therefore,
do  not  take  care  of  level  to  level  variations. Inspite  of  this
$C_{rel}$  is  quite  close to  the  value of  $C$ calculated  by  minimum
deviation  method. Among the  electronic  states  which  were  studied
by  Wei Hua \cite{r9}  the  values  of  $C_{rel}$  are  within $\pm 6\%$
of  the  values  of  the  $C$ except for $\rm Li_{2}\rm X^{1} \Sigma_{g}^{+3},
\rm H_{2}\rm X^{1}\Sigma_{g}^{+}, \rm CO X^{1}\Sigma_{g}^{+} \rm{and}
\rm XeO d^{1}\Sigma^{+}$  where the deviations are, respectively,
$11\%$, $15.9\%$, $24\%$ and  $24\%$. These deviations
can  be  understood  from  the  fact  that the  molecular  constants
vary  in going  to  higher  energy  levels. In  $\hbox{Li}_{2}$,
$v = 0$ to $4$  and  $v > 10$  levels  have  $\alpha_{e}$  values
$0.00704$  and  $0.0077$, respectively  (corresponding $\gamma_{e}$
values  are  also  different).  In  the case  of  those  electronic
states  for  which  complete  and  accurate  data  are  available,
$C_{rel}$   is  very  close  to  $C$; the  differences  may  be  attributed
to  the  fact  that  in  $C_{rel}$, finer  interactions  are  not  included.
\par
The  mean  square  deviations of  the  FP  potential  with
$C$ (i.e. $\Delta Y_{FPC}$) and $C_{rel}$(i.e.$\Delta Y_{FPC_{rel}}$) from
RKR  potential  curve have  been compared  in  Table 2. The  mean square
deviations  are within $\pm 5\%$  except  for  those  states  for
which $C_{rel}$ departs significantly  from  $C$.
\par
       A  stringent  condition  for  the  acceptance  of  a  potential
function  is  the  exact solution  of  the  corresponding  Schrodinger
wave  equation. As  discussed  by  Wei Hua \cite{r9}, FP  potential
provides  eigenvalues:
\be
E_{n} = \frac{D_{e}}{4} \left[ 2 + (Q^{2} + 1) - \frac{(\rho_{c}-\bar{n})^{2}}
{t^{2}} - \frac{(Q^{2}-1)^{2}t^{2}}{(\rho_{c}-\bar{n})^{2}} \right]
                                                           \label{27}
\ee
where
\bea
Q &=& \frac{1}{C}, t = \frac{2D_{e}}{\omega_{e}(1-C)},
\bar{n} = n + \frac{1}{2}, n = 0, 1, 2, 3; \nn \\
\rho_{c} &=& {\rm sign~~ of } (C){\rho} {\rm ~~ and ~~}
\rho = \left[ \frac{1}{4} + (Q^{2}-1)^{2}t^{2} \right]^{2}
                                                          \label{28}
\eea
The  mean  square  deviation  of  the  energy  values $(E_{cal})$  with
$C(\Delta E_{C})$ and  $C_{rel}(\Delta E_{C_{rel}})$  from  the  observed
levels ($E_{obs}$ i.e. $U_{RKR}$ values) have  been  calculated using  the
above  equation and  are  given  in  the  last  two  columns  of  Table  2.
Corresponding $\Delta E_{C}$ and $\Delta E_{C_{rel}}$  are very close to
each  other  supporting that $C_{rel}$ equals  $C$ within  the  accuracy
of  molecular  parameters and  can  be  computed  from  these (eqn. 18).
To  reinforce  the  above conclusions,  the  mean  square  deviations
viz. $\Delta Y_{j}$ and $\Delta E_{C_{rel}}$  from  the
RKR potential  have  been  compared  for  additional  ten  states
using  $C_{rel}$  in  the  FP  potential  function and  the  results
are  compiled  in  Table  3.
\par
      Taking  $C_{rel}$   as  the  correct  value of  $C$ parameters,
$\alpha_{e}$  and  $\omega_{e}x_{e}$ have  been calculated  using
equation 21  and  equation 22, for  37  electronic  states including
those studied  by  Wei Hua. The  FP  potential  with  $C_{rel}$   yields
$\alpha_{e}$  values  within $\pm 15 \%$  of the  corresponding  experimental
values. The  observed  variation  is  a  consequence  of uncertainty  in
the  values  of  the  molecular  parameters which, in turn,  determine  the
accuracy  of  $C_{rel}$. It  may,  however,  be  pointed  out  that
the  results  are  closer  to  the experimental  values  as compared  to the
ones  obtained  from  Lippincott's  potential function (eqn. 24)  which  is
claimed  to  be  the  best  potential  for  predicting  the  $\alpha_{e}$
values \cite{r4}. The  average  mean  deviations  for  FP, Morse, Varshni and
Lippincott's potentials  are  $7.2\%$,  $27.6\%$, $18.7\%$  and  $19.7\%$,
respectively,  establishing  the supremacy  of  FP  potential  over  other
three-parameter  potentials  in  predicting  the   $\alpha_{e}$ values.
\par
As  regards $\omega_{e}x_{e}$ the  average  mean  deviations  have  been
found  to  be $17.4\%$, $26.9\%$, $15.5\%$ and  $13.9\%$, respectively  for
FP, Morse,  Varshni  and  Lippincott's potential  functions. Obviously,
the  accuracy  is  not  as  good  as  for  $\alpha_{e}$. This  is perhaps
because  $C_{rel}$  itself  is  not  as  much  sensitive  to
$\omega_{e}x_{e}$. However,  the  results are  better  than  the  Morse
potential for  which, unlike  the  Varshni  and  Lippincott's potential
functions,  Schrodinger  equation  is  solvable. The  calculated  values
are slightly larger  than  the  experimental  values  for  almost   all
the  molecules.  A  graph between  $C_{rel}$   for  the  individual
level  and  $r_{max}$ (corresponding  to  v) (fig. 1) reveals that  $C$
increases
almost  exponentially  to  large  negative  values  at  higher  v. The
nature  of  the  curve  at  large  v  and  the  fact  that  the
factor $(1+C+C^{2})$ may  be approximated  to  $e^{C}(=1+C+\frac{C^{2}}{2})$
because $|C| < 1$, have  prompted  us  to  write $\omega_{e}x_{e}$
in the  light  of  equation  22, as
\be
\omega_{e}x_{e} = \frac{h}{8\pi^{2}c\mu}a^{2}e^{C}
                                                          \label{29}
\ee
The  use  of  this  relation  reduces  the  average  mean  deviation
to  $15.9\%$ (column 12 in Table 5)  which  is  comparable  to  the
value  obtained  from  Lippincott's  potential function.
\section{Conclusions}
The  constant $C$  has  been  expressed  in  terms  of  molecular  constants
implying that  the  FP  potential  manifests  the  contribution  of  vibration,
rotation, rotation-vibration  interaction  constants. Furthermore, the
potential  may  be  preferred  over  all the  known  three-parameter
potentials  for  the  prediction  of  the  rotation-vibration interaction
constant, $\alpha_{e}$  and  it  yields  the  value  of $\omega_{e}x_{e}$
as good  as  provided  by Lippincott's  function  which  is  known  to
be  the  best  analytical  function  for  its prediction.
\acknowledgments
The authors gratefully acknowledge  Prof. R.J.LeRoy for providing
the computer program used in the calculation of RKR potential. Both
the authors are thankful to Dr. Vishwamittar for fruitful discussions.
This work is supported by a grant from the DST, New  Delhi.

\listoftables
\begin{table}
\caption{Experimental Molecular Constants Used in this work. }
\vskip 0.4 cm
\tiny
\begin{tabular}{|l|l|l|c|c|c|c|c|c|c|c|c|c|l|l|l|l|l|}
\hline 
S.No. & Molecular & $ r_e $ & $D_e$ & $\omega_e$ & $\alpha_e$ 
& $ B_e $ & $\omega_e x_e$ & $\mu_e$ &Ref.\\
State & $ (A^{o}) $ &  $(cm^{-1})$ & $(cm^{-1})$ & $10^{3}(cm^{-1})$
  & $(cm^{-1})$ & $(cm^{-1})$ & & &
\\ \hline  \hline
1. & $Li_2 X^1 \Sigma_g^+$ & 2.6729 & 8516.780 & 351.430 & 7.040 & 0.6726 
& 2.6100 & 3.5080&17,14 \\ \hline
2. &$Na_2X^1\Sigma_g^+$ & 3.0788 & 6022.600 & 159.177 &$ 0.873$ & 0.1547
& 0.7254 & 11.4949&18,14 \\ \hline
3. &$K_2X^1\Sigma_g^+$ & 3.9244 &4440.000&92.405&0.212&0.0562&0.3276&19.4800 
&19\\ \hline
4.&$Cl_2X^1\Sigma_g^+$&1.9872 & 20276.440 & 559.751 & 1.516&0.2442&2.6943&
17.4844&20 \\ \hline
5.&$Cl_2B^3\Pi$ & 2.4311&3341.170 & 255.3800&2.511&0.1631&4.8000&17.48442&21
\\ \hline
6.&$I_2
XO_g^+$&2.6664&12547.335&214.520&0.113&$0.0373_7$&0.6079&63.4522&22,23
\\ \hline
7.&$ICl
X^1\Sigma^+$&2.3209&17557.600&384.275&0.532&$0.1142$&1.4920&27.4147
&24\\ \hline
8.&$ICl
A^3\Pi_1$&2.6850&3814.700&211.030&0.744&$0.0852_9$&2.1200&27.4147
&24,25\\ \hline
9.&$ICl A^3\Pi_2$&2.6651&4875.520&224.571&0.674&0.0865&1.8823&27.4147
&25\\ \hline
10.&$HF
X^1\Sigma^+$&0.9168&49384.000&4138.320&772.400&20.9557&89.8800&0.9571&26
\\ \hline
11.&$H_2
X^1\Sigma^+_g$&0.7416&38297.000&4401.265&3051.300&60.8477&120.6020&
0.5039&27 \\ \hline
12. & $CO X^1 \Sigma^+$  & 1.1283 & 90529.000 & 2169.813 & 17.504  &
1.93137 &
13.2883 & 6.8562&14 \\ \hline
13. & $XeO d^1 \Sigma^+$ & 2.8523 & 693.000 & 156.832 & 5.400 &$
0.1456$
& 9.8678  & 14.2327&28 \\ \hline
14. & $Cs_2
X^1\Sigma_g^+$&4.6480&3649.500&42.020&0.022&0.0117&0.0826&66.4527&30\\
\hline
15.&$Rb_2
X^1\Sigma_g^+$&4.2099&3950.000&57.7807&0.055&0.0224&0.1391&42.4559 &31\\
\hline
16. & $XeO b^1 \Sigma^+$ & 2.5480 & 461.000 & 113.636 & 14.593 & 0.1820
& 11.8410
& 14.2651&28 \\ \hline
17. & $Ar_2 CO_u^+$ &3.5960&465.800 & 66.820 & 2.500 & 0.0652 & 4.0000 &
19.9810&16\\ \hline
18. & $Ar_2 XO_g^+$ &3.7610 & 99.500 &30.6800 &3.641 &$ 0.0596_5$
&2.4200 &19.9810&16\\ \hline
19.&$ O_2 X^3\Sigma_g^-$ &1.2075 &42047.000 &1579.247 &15.466 &1.4456 &
11.5008
&7.9975&15 \\  \hline
20.&$O_2 b_1\Sigma_g^+$&1.2268 &28852.000 &1432.775 &18.198 &1.4004
&14.0065 &
7.9975&29 \\ \hline
21. & $O_2 A^3\Sigma_u^+$&1.5215 &6643.000 &815.648 &18.053&0.9105
&19.8513
&7.9975&15 \\ \hline
22.&$O_2
B^3\Sigma_u^-$&1.6042&8121.000&709.050&11.922&0.8189&10.6100&7.9975&15
\\ \hline
23.&$O_2^+ X^2\Pi_g$
&1.1171&54681.000&1905.335&18.970&1.6905&16.3040&7.9973
& 25 \\ \hline
24.& $NO X^2\Pi_{1/2}$
&$1.1507_7$&53323.758&1904.204&17.100&$1.6719_5$&14.0750&7.4664
&14\\ \hline
25.& $NO B^2\Pi$ & 1.4167
&26544.888&1037.200&12.000&1.0920&7.7000&7.4664 &14\\ \hline
26.&$N_2 X^1\Sigma_g^+$ &
1.0976&78742.304&2358.570&17.318&1.9982&14.3240&7.0015 &14 \\ \hline
27.&$N_2
A^3\Sigma^+_u$&1.2866&29772.23&1460.640&18.000&1.4546&13.8720&7.0015
&14\\ \hline
28.&$N_2 a^1\Pi_g$&
1.2203&48974.915&1694.208&17.930&1.6169&13.9490&7.0015&14 \\ \hline
29.&$N_2
B^3\Pi_g$&1.2126&39534.94&1733.390&17.910&1.6374&14.1220&7.0015&14 \\
\hline
30.& $OH X^2\Pi_i $
&0.9696&37308.074&3737.76&724.200&18.9108&84.8813&0.9481&14\\ \hline
31.& $OH
A^2\Sigma^+$&1.0121&20412.938&3178.860&786.800&17.3580&92.9170&0.9481
&14\\ \hline
32.& $ Br_2
X^1\Sigma_g^+$&2.2810&15900.307&325.321&0.318&0.0821&1.0774&39.4591&14
\\ \hline
33. &$ C_2
X^1\Sigma_g^+$&1.2425&50104.485&1854.710&17.650&1.8198&13.3400&6.0000&14\\
\hline
34.&$CO
d^3\Delta_i$&1.3696&28368.336&1171.940&17.820&1.3108&10.6350&6.8562&14\\
\hline
35.&$ CO A^1\Pi$&1.2353&25617.027&1518.240&23.530&1.6115&19.4000&6.8562
&14\\ \hline
36.&$ CO
e^3\Sigma^-$&1.3840&25391.113&1117.720&17.530&1.2836&10.6860&6.8562&14
\\ \hline
37.&$ CO
a^3\Sigma^+$&1.3523&34887.567&1228.600&18.920&1.3446&10.4680&6.8562&14
\\ \hline
\end{tabular}
\end{table}

\begin{table}
\caption{ Mean square deviations(eqn. 20) from the RKR curves for
the various potentials.}
\vskip 0.4 cm
\tiny
\begin{tabular}{|c|c|r|r|r|r|r|r|c|r|r|c|r|r|r|r|r|}
\hline 
S.No. & State & 
$C^{\rm a} $ & $C_{rel}$ & $\Delta Y_L$ & $\Delta Y_V$ & $\Delta Y_M$
& $\Delta Y_{FPC}$ & $\Delta Y_{FPC_{rel}}$ & $\Delta E_C$ & $\Delta
E_{C_{rel}}$ \\
& & & & ( \% ) & ( \% ) &( \% ) &( \% ) &( \% ) &( \% ) &( \% ) \\ 
\hline \hline 
1. & $Li_2 X^1 \Sigma_g^+$ & -0.1460 & -0.1298 & 9.879 & 8.498 & 9.952 &
2.842
& 2.985 & 2.576  & 2.707  \\ \hline          
2. &$Na_2X^1\Sigma_g^+$ & -0.2024 & -0.2031 & 18.721 & 16.058 & 21.372 &
2.028 & 2.027 & 2.878 & 2.872 \\ \hline
3. &$K_2X^1\Sigma_g^+$ &
-0.2780&-0.2694&10.448&9.155&13.622&2.034&2.062&1.374&1.409 \\ \hline
4.&$Cl_2X^1\Sigma_g^+$ & -0.1097&-0.0910&2.063 & 2.047 & 7.628 &
3.502&3.665&3.105&3.265 \\
\hline
5.&  $Cl_2B^3\Pi$ &
-0.1036&-0.1034&3.971&3.378&7.780&2.258&2.258&2.876&2.874 \\ \hline
6.&  $I_2 XO_g^+$         &
-0.1460&-0.1547&1.954&3.416&10.439&2.428&2.474&2.176&2.117 \\ \hline
7.&  $ICl X^1\Sigma^+$    &
-0.1000&-0.1020&1.822&1.828&6.842&3.298&3.300&3.306&2.989 \\ \hline
8.&  $ICl A^3\Pi_1$       &
-0.1680&-0.1780&3.036&5.452&12.181&1.104&1.244&0.910&0.925 \\ \hline
9.& $ICl A^3\Pi_2$      &
-0.1540&-0.1610&2.171&3.940&10.375&1.260&1.320&0.905&0.882   \\ \hline
10.& $HF X^1\Sigma^+$    &
0.1120&0.1210&4.170&4.313&6.313&3.211&3.245&2.816&2.921 \\ \hline
11.& $H_2 X^1\Sigma^+_g$  &
0.1510&0.1752&5.437&9.986&8.051&3.961&4.153&3.553&3.852 \\ \hline
12. & $CO X^1 \Sigma^+$   &   0.0370 & 0.0460 & 2.551 & 3.408 & 1.210 &
0.548 & 0.603 &
0.292 & 0.347 \\ \hline 
13. & $XeO d^1 \Sigma^+$ &  -0.0940 & -0.0683 & 7.478 & 6.192 & 7.570 &
5.183
& 5.398 & 5.140 & 4.957 \\ \hline
14. & $Cs_2
X^1\Sigma_g^+$&-0.2949&-0.3114&16.507&14.965&22.449&1.889&2.100&1.601&1.512\\
\hline
15.&$Rb_2
X^1\Sigma_g^+$&-0.2890&-0.2898&10.444&9.271&14.103&2.208&2.207&1.446&1.443\\
\hline
\hline
&Average& &&6.710&6.794&10.659&2.517&2.802&2.348&2.338\\ \hline
\end{tabular}
\vskip 0.4 cm
{\rm $^a$}The values are slightly different from those obtained by
Wei Hua \cite{r9},
possibly because  of reduced mass which we have taken from ref. 14
for all the molecules.
\end{table}

\begin{table}
\caption{Mean square deviations from the RKR curves for the various
potentials.}
\vskip 0.4 cm
\tiny
\begin{tabular}{|l|l|l|l|l|l|l|l|l|l|l|l|l|l|l|l|l|l|}
\hline 
S.No. & State & $C_F$ & $\Delta Y_L$ & $\Delta Y_V$ & $\Delta Y_M$ 
& $ \Delta Y_{FPC_{rel}}$ &  $ \Delta E_{C_{rel}}$&$ \ E_{max}$
&Ref.for\\
& & & ( \% ) & ( \% ) &( \% ) &( \% ) &( \% )&$\ D_e$&RKR Potential \\ 
\hline \hline 
1. & $XeO b^1 \Sigma^+$  & 0.5410 & 20.478 & 18.962 & 15.985 &  5.168 
& 2.984&0.6582&28 \\ \hline
2.&$Ar_2 CO_u^+$&0.1105&17.744&16.530&14.136&13.406&10.313&0.7022&16\\
\hline
3.&$Ar_2 XO_g^+$&-0.0223&7.302&5.967&6.881&6.038&2.015&0.9484&16\\
\hline
4.&$O_2 X^3\Sigma_g^-$&-0.0242&1.867&1.998&3.244&2.914&1.745&0.7064&15\\
\hline
5.&$O_2 b^1\Sigma_g^+$&-0.0530&0.419&0.484&1.347&0.851&0.381&0.3870&{\rm
(b)}\\
\hline
6.&$O_2
A^3\Sigma_u^+$&-0.2372&8.994&11.166&18.781&3.758&2.805&0.9849&15\\ \hline
7.&$O_2 B^3\Sigma_u^-$&-0.2500&4.612&5.097&11.171&2.612&0.927&0.9031&15
\\ \hline
8.&$O_2^+ X^2\Pi_g$&0.0036&0.968&1.220&0.110&0.094&0.102&0.3330&15\\
\hline
9.&$ CO a^3\Sigma^+$&0.0169&7.654&8.274&5.998&5.951&3.820&0.7664&{\rm
(b)}\\
\hline
10.&$N_2 X^1\Sigma_g^+$
&-0.0561&2.459&2.095&5.573&4.049&2.787&0.9937&{\rm (b)}\\ \hline
\hline
&Average& &7.25&7.18&8.32&4.48&2.79 & &\\ \hline
\end{tabular}
\vskip 0.4 cm
{\rm $^{(b)}$}RKR potential is calculated using LeRoy's computer
program \cite{r32}.
\end{table}

\begin{table}
\caption{Comparison of experimental values of $\alpha_e$ with calculated
values and the mean deviations
[$ \Delta X $=($\alpha_e(exptl)$-$\alpha_e(i)$)/($\alpha_e(exptl)$ ] ;
where $\alpha_e (i)$ stands for, $\alpha_e(FPC_{rel})$, FP;
$\alpha_e(M)$, Morse; $\alpha_e(V)$, Varshni;
$\alpha_e(L)$, Lippincott for different electronic states.}
\vskip 0.4 cm
\tiny
\begin{tabular}{|c|c|r|r|r|r|r|r|c|c|c|c|c|}
\hline 
S.No. & State &$
C_{rel}$&$\alpha_e(exptl)$&$\alpha_e(FPC_{rel})$&$\alpha_e(M)$&$\alpha_e(V)$&
$\alpha_e(L)$&$\Delta X_{FPC_{rel}}$&$\Delta X_M$&$\Delta X_V$&$\Delta X_L$\\ 
& & & $(10^{3} cm^{-1})$ &$(10^{3} cm^{-1})$& $(10^{3} cm^{-1})$&
$(10^{3} cm^{-1})$&
$(10^{3} cm^{-1})$& (\%) & (\%) & (\%) & (\%)  \\ \hline \hline 

1. & $Li_2  X^1 \Sigma_g^+$ & -0.1298&  7.040&  7.875&
10.206&9.137&9.097 &11.9&45.0&29.8&29.2
\\ \hline
2. &$Na_2  X^1
\Sigma_g^+$&-0.2031&0.873&0.972&1.449&1.239&1.282&11.4&66.1&42.0&46.9\\
\hline
3. &$K_2 X^1\Sigma_g^+$
&-0.2694&0.212&0.233&0.394&0.329&0.346&9.9&86.0&55.4&63.5 \\ \hline
4.&$Cl_2
X^1\Sigma_g$&-0.0910&1.516&1.672&1.903&1.585&1.655&10.3&25.5&4.6&9.1
\\ \hline
5.&  $Cl_2B^3\Pi$
&-0.1034&2.511&2.439&2.793&2.396&2.412&2.8&11.2&4.6&3.9\\ \hline
6.&  $I_2 XO_g^+$
&-0.1547&0.113&0.125&0.154&0.131&0.134&10.1&36.5&15.9&18.1\\ \hline
7.&  $ICl X^1\Sigma^+$
&-0.1020&0.532&0.581&0.670&0.561&0.582&9.2&26.0&5.5&9.3\\ \hline
8.&  $ICl A^3\Pi_1$ &
-0.1780&0.744&0.788&1.000&0.867&0.865&5.8&34.8&16.5&16.3\\ \hline
9.& $ICl A^3\Pi_2$ &
-0.1610&0.674&0.717&0.893&0.767&0.772&6.4&32.5&13.7&14.4\\ \hline
10.& $HF X^1\Sigma^+$
&0.1210&772.400&801.900&658.200&647.600&593.900&3.8&14.8&16.1&23.1 \\
\hline
11.& $H_2 X^1\Sigma^+_g$ &
0.1752&3051.300&3502.874&2229.439&4183.994&2161.944&14.8&26.9&37.1&29.1\\
\hline
12. & $CO X^1 \Sigma^+$
&0.0460&17.504&17.600&16.440&14.080&14.547&0.5&6.0&19.5&16.9\\ \hline
13. & $XeO d^1 \Sigma^+$
&-0.0683&5.400&5.088&5.520&4.932&4.745&5.8&2.2&8.9&12.1 \\ \hline
14. & $Cs_2
X^1\Sigma_g^+$&-0.3114&0.022&0.0238&0.0435&0.0361&0.0381&8.8&98.6&64.7&73.9\\
\hline
15.&$Rb_2
X^1\Sigma_g^+$&-0.2898&0.055&0.062&0.108&0.0897&0.0947&11.3&95.3&62.4&71.3\\
\hline
16. & $XeO b^1 \Sigma^+$
&0.5410&14.593&14.900&9.090&7.910&7.842&2.0&37.6&45.7&46.2 \\ \hline
17.&$Ar_2
CO_u^+$&0.1105&2.500&2.189&1.930&1.670&1.667&12.4&22.6&32.8&33.3 \\
\hline
18.&$Ar_2
XO_g^+$&-0.0223&3.641&3.580&3.685&3.210&3.176&1.6&1.2&11.8&12.7\\
\hline
19.&$O_2 X^3\Sigma_g^-$&
-0.0242&15.466&16.870&17.480&14.500&15.311&8.6&13.0&6.2&0.9\\ \hline
20.&$O_2 b^1\Sigma_g^+$&
-0.0530&18.198&19.500&21.052&17.449&18.364&7.1&15.6&4.1&0.9\\ \hline
21.&$O_2
A^3\Sigma_u^+$&-0.2372&18.053&18.290&25.870&22.100&22.360&1.3&43.3&22.4&23.9\\
\hline
22.&$O_2 B^3\Sigma_u^-
$&-0.2500&11.922&12.810&18.990&15.920&16.478&7.4&59.3&33.6&38.2\\
\hline
23.&$O_2^+ X^2\Pi_g$&
-0.0036&18.970&19.130&18.880&15.690&16.680&1.7&0.5&17.4&12.2\\ \hline
24. &$NO X^2\Pi_1$ &
-0.0352&17.100&18.180&19.000&15.770&16.600&5.8&11.1&7.7&2.6\\ \hline
25.& $NO B^2\Pi$ &
-0.0371&12.000&13.030&13.802&11.500&12.120&8.3&15.0&4.1&1.0\\ \hline
26.& $N_2 X^1\Sigma_g^+$ &
-0.0561&17.318&18.478&20.030&16.710&17.598&6.6&15.6&3.5&1.6\\ \hline
27.& $N_2 A^3\Sigma_u^+$
&-0.0836&18.000&19.250&21.808&18.070&19.030&6.7&21.1&0.4&5.7\\ \hline
28.& $N_2 a^1\Pi_g$ &
-0.0157&17.930&18.170&18.610&15.500&16.330&1.0&3.7&13.5&8.9\\ \hline
29.& $N_2 B^3\Pi_g$
&-0.0915&17.910&19.400&22.300&18.490&19.500&8.0&24.6&3.2&8.9 \\ \hline
30.& $OH X^2\Pi_i$ &
0.0581&724.200&777.000&703.110&645.100&628.900&7.2&2.9&10.9&13.2\\
\hline
31.& $OH A^2\Sigma^+$&
-0.0518&786.800&949.600&806.990&650.600&838.700&15.8&20.7&2.5&6.6 \\
\hline
32.& $Br_2 X^1\Sigma^+_g$
&-0.1431&0.318&0.350&0.430&0.360&0.377&11.4&36.6&14.9&18.4\\ \hline
33.& $C_2 X ^1\Sigma_g^+$ &
-0.1013&17.650&18.850&22.180&18.440&19.459&6.5&25.6&4.5&10.2\\ \hline
34.& $CO d^3 \Delta_i$&
0.0398&17.820&18.990&17.930&14.920&15.735&6.6&0.6&16.2&11.7\\ \hline
35.&$CO A^1\Pi $&
-0.1072&23.530&23.970&28.080&23.300&24.450&4.8&19.9&0.9&3.9 \\ \hline
36.&$CO e^3\Sigma^-$&
-0.0014&17.530&18.530&15.400&12.500&16.249&5.4&5.7&12.1&7.3 \\ \hline
37.&$ CO a^3
\Sigma^+$&0.0169&18.920&16.664&16.212&13.608&14.271&12.1&14.3&28.1&24.6
\\ \hline
\hline
&Average& &  & & &&&7.2&27.6&18.7&19.7\\ \hline
\end{tabular}
\end{table}

\begin{table}
\caption{Comparison of experimental values of $\omega_e x_e$ with
calculated values and the mean deviations
[$ \Delta X $=($\omega_e x_e(exptl)$-$\omega_e x_e(i)$)/($\omega_e
x_e(exptl)$ ] ; where $\omega_e x_e (i)$ stands for $\omega_e
x_e(FC_{rel})$, FP;
$\omega_e x_e(M)$, Morse; $\omega_e x_e(V)$, Varshni; $\omega_e x_e(L)$,
Lippincott; $\omega_e x_e(exp.)$ using equation 29 for the
corresponding electronic states given in table 3.}
\vskip 0.4 cm
\tiny
\begin{tabular}{|c|c|c|r|r|r|r|c|c|c|c|c|c|c|c|c|c|c|c|}
\hline 
S.No. & $\omega_ex_e(exptl.)$&$\omega_ex_e
FPC_{rel}$&$\omega_ex_e(M)$&$\omega_ex_e(V)$
&$\omega_ex_e(L)$ & $\omega_e x_e(exp)$ & $\Delta X_{FPC_{rel}}$ &
$\Delta X_M$ & $\Delta X_V$ & $\Delta X_L$ & $\Delta X_exp$ \\
&$ cm^{-1}$  &$ cm^{-1}$ &$ cm^{-1}$ &$ cm^{-1}$ &$ cm^{-1}$ &$ cm^{-1}$
& (\%) & (\%)& (\%) & (\%)&(\%)  \\ \hline \hline
1.  &2.6100&3.2150&3.6250&3.9500&2.9708&3.1835&23.1&38.8&16.1&13.8&21.9
\\ \hline
2.  &0.7254&0.8816&1.0518&0.8640&0.8468&0.8585&21.5&45.0&19.1&16.7&18.3
\\ \hline
3.
&0.3276&0.3859&0.4808&0.3918&0.3814&0.3700&17.8&46.8&19.6&16.4&12.0\\ \hline
4.&2.6943&3.5437&3.8631&3.1681&2.9890&3.5272&31.5&43.4&17.6&10.9&30.9
\\ \hline
5.&4.8000&4.4261&4.8800&4.0982&3.7199&4.3991&7.8&1.7&14.6&22.5&8.3\\
\hline
6.&0.6079&0.7973&0.9169&0.7637&0.7016&0.7860&31.2&50.8&25.6&15.4&29.3\\
\hline
7.&1.4920&1.9104&2.1026&1.7328&1.6202&1.8992&28.01&40.9&16.1&8.6&27.3\\
\hline
8.
&2.1210&2.4911&2.9185&2.466&2.2206&2.4423&17.4&37.6&16.3&4.7&15.5\\ \hline
9.
&1.8823&2.2354&2.5859&2.1712&1.9709&2.2002&18.8&37.4&15.3&4.7&16.9\\ \hline
10.
&89.8800&98.4446&86.6965&75.3269&72.8732&97.8364&9.5&3.5&16.2&18.9&8.8\\
\hline
11.
&120.6020&152.4790&126.4533&148.9109&117.6726&150.6605&26.4&4.9&23.5&2.4&24.9\\
\hline
12.
&13.2883&13.6250&13.0016&13.6854&10.4741&13.6117&2.5&2.2&19.6&21.2&2.4\\
\hline
13.
&9.8678&8.3081&8.8731&7.7068&6.7091&8.2869&15.8&10.0&21.9&32.0&16.0\\ \hline
14.&0.0826&0.0950&0.1209&0.09848&0.09510&0.8858&15.1&46.5&19.3&15.2&7.3
\\ \hline
15.&0.1391&0.1678&0.2113&0.1721&0.1668&0.1581&20.6&51.9&23.7&19.9&13.6
\\ \hline
16.
&11.8410&12.8083&7.0028&5.9355&5.3070&11.9983&8.2&40.9&49.8&55.1&1.3\\ \hline
17.&4.0000&2.6900&2.3960&2.0304&1.8202&2.6742&32.8&44.1&49.2&54.5&33.1\\
\hline
18. &2.4200&2.3130&2.3649&2.0126&1.7960&2.3124&4.4&2.3&16.8&25.7&4.4\\
\hline
19.&11.5008&14.4758&14.8287&12.0728&11.6615&14.4715&25.9&28.9&4.9&1.4&25.8\\
\hline
20.
&14.0065&16.8908&17.7877&14.5104&13.8627&16.8654&20.6&26.9&3.6&1.0&20.4\\
\hline
21.&19.8513&20.5039&25.0369&20.9516&19.1156&19.7470&3.3&26.1&5.5&3.7&0.5
\\ \hline
22.&10.6100&12.5700&15.4769&12.758&11.9120&12.0516&18.4&45.8&20.2&12.3&13.6\\
\hline
23.&16.3040&16.6684&16.5940&13.5247&13.0900&16.6683&2.2&1.8&17.0&19.7&2.2\\
\hline
24.&14.0750&16.1025&16.9990&13.5732&13.1284&16.0920&14.4&20.7&3.6&6.7&14.3\\
\hline
25.&7.7000&9.4823&10.1317&8.0125&7.7847&9.4755&23.1&31.6&4.1&1.1&23.1\\
\hline
26.&14.3240&16.7209&17.6615&14.3895&13.9912&16.6926&16.7&23.3&0.5&2.3&16.5
\\ \hline
27.&13.8720&16.5410&17.9149&14.6094&13.9800&16.4767&19.2&29.1&5.3&0.8&18.7\\
\hline
28.&13.9490&14.4239&14.6521&11.9364&11.5933&14.4221&3.4&5.0&14.4&16.9&3.4\\
\hline
29.&14.1220&17.4174&18.9990&15.4808&14.8614&17.3355&23.3&34.5&9.6&5.2&22.7\\
\hline
30.&84.8813&99.3482&93.6181&79.0097&77.2874&99.1933&17.0&10.3&6.9&8.9&16.8\\
\hline
31.&92.9170&117.6646&123.7592&101.4363&99.3160&117.4958&26.6&33.1&9.2&6.9&26.4
\\ \hline
32.
&1.0774&1.4596&1.6640&1.3753&1.2785&1.4418&35.4&54.4&27.7&18.7&33.6\\ \hline
33.
&13.3400&15.5980&17.1638&13.9766&13.5530&15.5071&16.9&28.6&4.8&1.6&16.2 \\
\hline
34.
&10.6350&12.6026&12.1036&9.8585&9.5680&12.5931&18.0&13.8&7.3&10.1&18.4\\
\hline
35.&19.4000&20.3393&22.4953&18.3891&17.4757&20.2056&1.7&15.9&5.2&9.9&4.1
\\ \hline
36.&10.6860&12.2812&12.3005&10.0159&9.7050&12.2812&14.9&15.1&6.3&9.2&14.9\\
\hline
37.&10.4680&11.0015&10.8165&8.8304&8.6160&11.0000&5.1&3.3&15.6&17.7&5.1\\
\hline
\hline
&Average& &  & & &&17.4&26.9&15.5&13.4&15.9\\ \hline
\end{tabular}
\end{table}

\end{document}